# Comment on “Observation of dark pulse in a dispersion-managed fiber ring laser”

Han Zhang, Dingyuan Tang*, Luming Zhao, and Xuan Wu

*School of Electrical and Electronic Engineering, Nanyang Technological University, Singapore 639798*

* Corresponding author: edytang@ntu.edu.sg

A recent communication [Opt. Commun. doi:10.1016/j.optcom.2010.06.076 (2010)] presents experimental results in which dark pulses are observed in a dispersion-managed (DM) net-anomalous dispersion fiber laser. Disagreement on the formation mechanism proposed in this communication, we would like to indicate a more accurate explanation in order to clarify some potential misunderstanding on dark pulses in fiber lasers.

In a recent communication by Yin et al., although operated in anomalous dispersion regime, dark pulses were still observed and interpreted by the balance of linear and nonlinear inter-modulation effects [1]. Here, we would like to briefly summarize the two types of formation mechanisms for dark pulses and point out that Yin's explanation is less supportive. One is nonlinear Schrödinger equation (NLSE) type dark pulse in [2] while the other is the domain wall (DW) type dark pulse [3, 4]. While a NLSE type dark pulse was observed in fiber lasers of normal dispersion, a DW type dark pulse was obtained in fiber lasers of either anomalous or normal dispersion. Based on our early contributions, we could distinguish these two types of dark pulses through: 1) NLSE type dark pulse had a single-peak profile, as shown in the [2] while DW type dark pulse had multi-peak structure, as shown in [3, 4]; 2) DW type dark pulse had lower onset threshold (~ 100-mw pump in [3, 4]) than NLSE type dark pulse (~ 2-W pump in [2]); 3) DW type dark pulse had broader pulse width (nanosecond scale in [4]) while NLSE type dark pulse had narrower pulse width (pico-second in [2]). In combination with this communication [1], we concluded that what Yin reported were actually DW type dark pulses. The emergence of intensity dips originated from the mutual coupling of different wavelength beams, leading to the emergence of topological defects in temporal domain. Obviously, this process is dispersion independent. The dispersion induced walk off between different optical waves does not require additional compensation but affects the temporally topological defect, corresponding to dark pulse width. Whereas, according to their explanation, it seemed that dispersion managed cavity was a prerequisite to form DW type dark pulses in that the walk off of pulses in anomalous dispersion fiber was counteracted by the normal dispersion in order to reach the balance of linear and nonlinear inter-modulation effects. This contradicted with what we experimentally observed in [4].